\documentclass[prb,twocolumn]{revtex4-1}

\usepackage{amsmath}
\usepackage{amsfonts}
\usepackage{graphicx}
\usepackage{color}
\usepackage{verbatim}

\begin{document}

\title{Tunneling Anisotropic Thermopower and Seebeck Effects in Magnetic Tunnel Junctions}

\author{Carlos L\'opez-Mon\'is, Alex Matos-Abiague and Jaroslav Fabian}

\affiliation{Institut f\"ur Theoretische Physik, Universit\"at Regensburg, 93040 Regensburg, Germany}

\date{\today}


\begin{abstract}
The Tunneling Anisotropic Magneto-Thermopower (TAMT) and the Tunneling Anisotropic Spin-Seebeck (TASS) effects are studied for a magnetic tunnel junction (MTJ) composed of a ferromagnetic electrode, a zinc-blende semiconductor and a normal metal. We develop a theoretical model for describing the dependence of a thermally induced tunneling current across the MTJ on the in-plane orientation of the magnetization in the ferromagnetic layer. The model accounts for the specific Bychkov-Rashba and Dresselhaus spin-orbit interactions present in these systems, which are responsible for the $C_{2v}$ symmetry we find in the TAMT and the TASS.
\end{abstract}


\maketitle



\section{Introduction}

Magneto-thermoelectric phenomena have been thoroughly studied for a long time.~\cite{Barnard,Ziman} More than twenty years ago Johnson and Silsbee investigated thermoelectric magnetization transport across ferromagnetic-paramagnetic interfaces.~\cite{Johnson_PRB_1987} However, only during the last lustrum thermoelectric transport has been able to account for the electronic spin degree of freedom,~\cite{Zutic_RMP_2004,ACTA} mainly due to the the discovery of the spin Seebeck effect.~\cite{Uchida_Nature_2008,Uchida_NatMat_2010,Jaworski_NatMat_2010} This has given birth to the field of spin caloritronics,~\cite{Bauer_SSC_2010,Bauer_NatMat_2012} which covers the non-equilibrium phenomena related with heat, charge and {\it spin} transport in small magnetic structures.

A three layer magnetic tunnel junction (MTJ) is a heterostructure consisting of a ferromagnetic conductor, a tunnel barrier, and a ferromagnetic or normal conductor, which are all grown one on top of the other. In the growth direction, the dimension of the MTJ layers is on the nanometric scale, thereby quantum effects have to be considered. Additionally, electric fields, spin voltages and temperature gradients are used to probe their physical properties. These forces create measurable currents that transport charge, spin and/or heat across the MTJ, making them promising systems for investigating spin caloritronic phenomena.~\cite{Walter_NatMat_2011,Breton_Nature_2011,Liebing_PRL_2011,Lin_NatCom_2012,Flipse_NatNano_2012,Bauer_NatMat_2012,Jeon_SciRep_2012,Ni_SciRep_2013,Vera-Marun_arXiv_2013}

Experimentally, it has been observed that the resistance of a MTJ depends on the magnetization orientation with respect to the crystallographic axes of the ferromagnetic layers.~\cite{Tanaka_PRL_2001,*Higo_APL_2001,Chiba_PhysE_2004,Ruster_PRL_2005,Saito_PRL_2005,Gao_PRL_2007} This effect is known as tunneling anisotropic magnetoresistance (TAMR). Surprisingly, TAMR has also been observed in MTJs with a single magnetic layer.~\cite{Gould_PRL_2004,Moser_PRL_2007,Liu_NanoLett_2008} However, the behavior of the TAMR varies depending on (i) the specific composition of the MTJ, (ii) if it has one or two magnetic electrodes, and (iii) whether the magnetization is rotated within a plane perpendicular to the ferromagnetic layer ({\it out-of-plane} TAMR) or in the plane of the ferromagnetic layer ({\it in-plane} TAMR). MTJs can be made of several kinds of materials. (Ga,Mn)As magnetic semiconductors are the most commonly used ferromagnetic electrodes,~\cite{Tanaka_PRL_2001,*Higo_APL_2001,Chiba_PhysE_2004,Gould_PRL_2004,Ruster_PRL_2005,Saito_PRL_2005} however TAMR has also been observed in MTJs using Fe~[\onlinecite{Moser_PRL_2007}] and Co~[\onlinecite{Liu_NanoLett_2008}] transition metals (or CoFe~[\onlinecite{Gao_PRL_2007}] alloys). Regarding the tunnel barrier, TAMR has been seen with both semiconductor barriers (e.g., GaAs~[\onlinecite{Chiba_PhysE_2004,Ruster_PRL_2005,Moser_PRL_2007}], AlAs~[\onlinecite{Tanaka_PRL_2001,*Higo_APL_2001}] or ZnSe~[\onlinecite{Saito_PRL_2005}]) and insulator barriers (e.g., Al$_2$O$_3$~[\onlinecite{Gao_PRL_2007,Gould_PRL_2004,Liu_NanoLett_2008}] or MgO~[\onlinecite{Gao_PRL_2007}]). And for single magnetic layer MTJs, Au is the most commonly used non magnetic electrode.~\cite{Gould_PRL_2004,Moser_PRL_2007,Liu_NanoLett_2008} TAMR phenomena are not restricted to MTJs; they have been observed, for example, in single Co atoms.~\cite{Neel_PRL_2013} Recently, anisotropic electric spin-injection has also been achieved for ferromagnetic-Silicon interfaces.~\cite{Sharma_PRB_2012}

The anisotropy of the tunneling magnetoresistance is due to spin-orbit interaction (SOI).~\cite{Brey_APL_2004} However, the role SOI plays depends on the specific composition of the MTJ. In (Ga,Mn)As based MTJs, the origin of the TAMR seems to be the anisotropic density of states (DOS) of the ferromagnetic semiconductor with respect to magnetization, due to its strong SOI in the valence band, combined also with uniaxial strain effects.~\cite{Gould_PRL_2004,Ruster_PRL_2005,Saito_PRL_2005,Sankowski_PRB_2007} For MTJs with transition metal electrodes and an insulating barrier, the Rashba shift of the interface resonant states can produce TAMR,~\cite{Gao_PRL_2007,Khan_JPCM_2008,Chantis_PRL_2007} although the anisotropy in the DOS also might induce TAMR.~\cite{Liu_NanoLett_2008} Finally, for MTJs fabricated with transition metal electrodes and a zinc-blende type semiconductor barrier, TAMR is proposed to be due to interface Bychkov-Rashba (BR) and Dresselhaus (D) SOI.~\cite{Moser_PRL_2007,Matos-Abiague_PRB_2009,Matos-Abiague_PRB_2009-2,Wimmer_PRB_2009}

Recently, an experiment has been performed in the all-semiconductor MTJ (Ga,Mn)As/GaAs/Si:GaAs, in which electric current was driven by a thermal gradient ---instead of an electric field--- in order to study the in-plane tunneling anisotropic magneto-thermopower effect (TAMT).~\cite{Naydenova_PRL_2011} The thermopower measures the voltage (or current, for the case of closed circuits) induced by a temperature gradient. Hence, for the case of MTJs this quantity is dubbed tunneling magneto-thermopower. Naydenova {\it et al.}~\cite{Naydenova_PRL_2011} have measured an anisotropic dependence of the thermopower on the magnetization orientation of the ferromagnetic electrode with respect to a reference crystallographic axis. As in the TAMR case, the anisotropy is likely due to the effect of the strong SOI on the DOS of the ferromagnetic semiconductor. Theoretically, the anisotropy of the thermopower induced by SOI has been investigated for a normal-metal/helical-multiferroic/ferromagnetic MTJ.~\cite{Jia_APL_2011}

In this paper, we investigate TAMT and the tunneling anisotropic spin-Seebeck effect (TASS). The former describes the anisotropy of the thermopower (also called Seebeck coefficient), while the latter effect describes the anisotropy of the spin-Seebeck coefficient. We focus on MTJs composed of transition metal electrodes, a zinc-blende semiconductor barrier and with a single ferromagnetic layer. The calculation is based on the model introduced in Refs.~[\onlinecite{Moser_PRL_2007,Matos-Abiague_PRB_2009}], which accounts for the BR and D SOIs that are likely to be the cause of the anisotropy for these kind of MTJs. The main result of this work is the characteristic $C_{2v}$ symmetry found for the TAMS and the TASS, similar to the one observed for the TAMR in the same system.~\cite{Moser_PRL_2007}

The paper is organized as follows. The theory is presented in Sec.~\ref{theory-sec}: The TAMT and the TASS are defined in Sec.~\ref{def-sec}, the tunneling current is computed in Sec.~\ref{current-sec}, and the model for describing the MTJ is presented in Sec.~\ref{model-sec}, namely, the Hamiltonian (Sec.~\ref{H-sec}), the tunneling states (Sec.~\ref{PS-sec}) and the transmission probability (Sec.~\ref{T-sec}). The results are presented in Sec.~\ref{results-sec}, where a phenomenological model is used for describing qualitatively the results (Sec.~\ref{phe-sec}). Finally, a summary is given in Sec.~\ref{summary-sec}.


\section{Theory} \label{theory-sec}

\subsection{Definitions} \label{def-sec}

\begin{table}
  \begin{tabular}{cccccc}
    \hline \hline
    Ratio & $\nabla \mu_c$ & $\nabla \mu_s$ & ${\bf I}_c$ & ${\bf I}_s$ & Material Property \\ \hline
    $-\nabla \mu_c/e\nabla T$ & $\checkmark$ & $0$ & $0$ & $\checkmark$ & $S$ \\ 
    $-\nabla \mu_c/e\nabla T$ & $\checkmark$ & $\checkmark$ & $0$ & $0$ & $S_+/2$ \\
    $-\nabla \mu_s/2e\nabla T$ & $\checkmark$ & $\checkmark$ & $0$ & $0$ & $S_-/2$ \\ 
    $-\nabla \mu_s/2e\nabla T$ & $0$ & $\checkmark$ & $\checkmark$ & $0$ & $S_s$ \\
    \hline \hline
  \end{tabular}
  \caption{The material property described by the ratio between an electrochemical potential (spin accumulation) gradient, $\nabla \mu_c$ ($\nabla \mu_s$), and a temperature gradient, $\nabla T$, depends on $\nabla \mu_s$ ($\nabla \mu_c$), ${\bf I}_c$ and ${\bf I}_s$. The material properties $S$, $S_+$, $S_-$ and $S_s$ are defined in Eqs.~\eqref{Sc},~\eqref{S+},~\eqref{S-} and~\eqref{Ss}, respectively. The check mark symbol (\checkmark) means that the corresponding quantity is finite.}
  \label{table-b.c.}
\end{table}

In general, the induced current across a MTJ associated to the spin-$\sigma$ channel, $I_{\sigma}$, is given by the constitutive equation~\cite{Gravier_PRB_2006,Jansen_PRB_2012}
\begin{equation} \label{Isigma-grl}
  {\bf I}_{\sigma} = -G_{\sigma} \left( \frac{\nabla \mu_{\sigma}}{e} + S_{\sigma} \nabla T \right)
\end{equation}
where $\sigma = \uparrow, \downarrow$. $G_{\sigma}$ and $S_{\sigma}$ are the conductance and the Seebeck coefficient of the spin-$\sigma$ channel, respectively, $\mu_{\sigma}$ is the spin-$\sigma$ electrochemical potential and $\nabla T$ an applied thermal gradient. Using Eq.~\eqref{Isigma-grl} we can write,
\begin{subequations}
  \begin{eqnarray}
    {\bf I}_c & = & -G \left( \frac{\nabla \mu_c}{e} + P\frac{\nabla \mu_s}{2e} + S \nabla T \right), \label{Ic} \\
    {\bf I}_s & = & -G \left( P\frac{\nabla \mu_c}{e} + \frac{\nabla \mu_s}{2e} + S_s \nabla T \right), \label{Is}
  \end{eqnarray}
\end{subequations}
where ${\bf I}_{c/s} = {\bf I}_{\uparrow} \pm {\bf I}_{\downarrow}$ is the charge/spin current, $\mu_c = (\mu_{\uparrow} + \mu_{\downarrow})/2$ and $\mu_s = \mu_{\uparrow} - \mu_{\downarrow}$ are the charge electrochemical potential and the spin accumulation, respectively, and $P = G_s/G$, where $G = G_{\uparrow} + G_{\downarrow}$ is the conductance and $G_s = G_{\uparrow} - G_{\downarrow}$ the spin conductance. Furthermore, 
\begin{equation}
  S = \frac{1}{G} \left( G_{\uparrow} S_{\uparrow} + G_{\downarrow} S_{\downarrow} \right), \label{Sc}
\end{equation}
is the commonly referred to as the {\it thermopower} (or Seebeck coefficient),~\cite{Slachter_NatPhy_2010,Czerner_PRB_2011,Bauer_NatMat_2012} and
\begin{equation}
  S_s = \frac{1}{G} \left( G_{\uparrow} S_{\uparrow} - G_{\downarrow} S_{\downarrow} \right), \label{Ss}
\end{equation}
which in the following we shall name {\it effective spin-Seebeck coefficient}. Additionally, although the quantities
\begin{subequations}
  \begin{eqnarray}
    S_+ & = & S_{\uparrow} + S_{\downarrow}, \label{S+} \\
    S_- & = & S_{\uparrow} - S_{\downarrow}, \label{S-}
  \end{eqnarray}
\end{subequations}
appear neither in Eq.~\eqref{Isigma-grl} nor in Eqs.~\eqref{Ic} and~\eqref{Is}, they as well are relevant material properties. In the literature, $S_-$ is commonly referred to as the {\it spin-Seebeck coefficient},~\cite{Slachter_NatPhy_2010,Jia_APL_2011,Czerner_PRB_2011,Flipse_NatNano_2012,Scharf_PRB_2012} while hereinafter we shall refer to $S_+$ as the {\it effective thermopower}. 

Table~\ref{table-b.c.} shows how the measurement of the aforementioned material properties depends on $\nabla \mu_c$, $\nabla \mu_s$, ${\bf I}_c$ and ${\bf I}_s$. For example, in an open circuit setup (${\bf I}_c = 0$), the ratio $\nabla \mu_s/\nabla T$ is related to the spin-Seebeck coefficient $S_-$ [Eq.~\eqref{S-}]; whereas the ratio $\nabla \mu_c/\nabla T$ will be related either to the thermopower, $S$ [Eq.~\eqref{Sc}] or to the effective thermopower $S_+$ [Eq.~\eqref{S+}], depending on whether there is a finite spin accumulation gradient or not. However, in a closed circuit setup (${\bf I}_c \ne 0$), Table~\ref{table-b.c.} shows that the ratio $\nabla \mu_s/\nabla T$ is now related to the effective spin-Seebeck coefficient $S_s$ [Eq.~\eqref{Ss}].

\begin{figure}
  \includegraphics[scale=0.35]{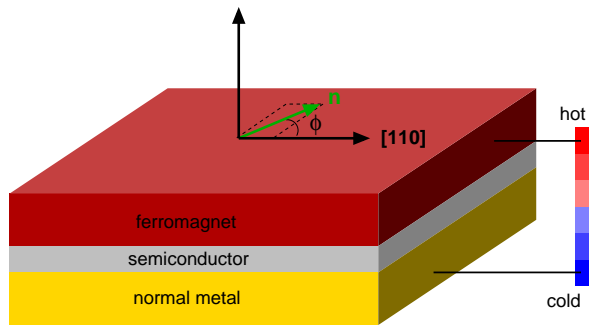}
  \caption{(Color online.) Scheme of a three layer magnetic tunnel junction. A thermally induced current tunnels across the semiconductor from the ferromagnet electrode into the normal metal. Spin-orbit interaction is responsible for an anisotropic dependence of the (effective) thermopower and the (effective) spin-Seebeck coefficient on the in-plane magnetization orientation ${\bf n}$ (green arrow) with respect to a reference crystallographic axis, which in the present case has been taken as the GaAs $[110]$ direction.}
  \label{fig-MTJ}
\end{figure}

The Tunneling Anisotropic Magneto-Thermopower, measures the relative dependence of the thermopower [Eq.~\eqref{Sc}] or the effective thermopower [Eq.~\eqref{S+}] on the in-plane magnetization orientation, hence, we have that
\begin{subequations}
  \begin{eqnarray}
    {\rm TAMT}(\phi) & = & \frac{S(0) - S(\phi)}{S(\phi)}, \quad {\rm or} \label{TAMS} \\
    {\rm TAMT}_{[+]}(\phi) & = & \frac{S_+(0) - S_+(\phi)}{S_+(\phi)}, \label{TAMS+}
  \end{eqnarray}
\end{subequations}
where $\phi$ is the angle spanned between the magnetization vector and a reference crystallographic axis $[x]$ in the ferromagnet layer (see Fig.~\ref{fig-MTJ}). Likewise, the Tunneling Anisotropic Spin-Seebeck ratio measures the relative dependence of the spin-Seebeck coefficient [Eq.~\eqref{S-}] or the effective spin-Seebeck coefficient [Eq.~\eqref{Ss}] on the in-plane magnetization orientation, hence, we have that
\begin{subequations}
\begin{eqnarray}
  {\rm TASS}_{[-]}(\phi) & = & \frac{S_-(0) - S_-(\phi)}{S_-(\phi)}, \quad {\rm or} \label{TASS-} \\
  {\rm TASS}(\phi) & = & \frac{S_s(0) - S_s(\phi)}{S_s(\phi)}.  \label{TASSs}
\end{eqnarray}
\end{subequations}

In the remaining, in order to compute the Seebeck coefficient $S_{\sigma}$ of the spin-$\sigma$ channel, we consider solely a temperature gradient as the driving force responsible for the tunneling current across the MTJ. Thus, Eq.~\eqref{Isigma-grl} reduces to
\begin{equation} \label{Isigma}
  {\bf I}_{\sigma} = -G_{\sigma} S_{\sigma} \nabla T.
\end{equation}


\subsection{Tunneling Current} \label{current-sec}

The net current that flows across the MTJ for the spin-$\sigma$ channel is~\cite{ACTA}
\begin{equation} \label{Isint}
  I_{\sigma} = \frac{1}{e} \int g_{\sigma}(E) [f_L(E) - f_R(E)] dE
\end{equation}
where $f_L(E)/f_R(E)$ is the Fermi-Dirac distribution of the left/right electrode,
\begin{equation} \label{GE}
  g_{\sigma}(E) = \frac{e^2}{h} \frac{1}{(2 \pi)^2} \int T_{\sigma}({\bf k}_{\|},E) d^2{\bf k}_{\|},
\end{equation}
and $T_{\sigma}({\bf k}_{\|},E)$ is the transmission probability of the spin-$\sigma$ channel,
namely, the probability of an electron with spin-$\sigma$ to tunnel through a potential barrier. The integrals in Eqs.~\eqref{Isint} and~\eqref{GE} are performed over the energy, $E$, and the transverse modes, ${\bf k}_{\|}$ (see next section).

In the linear response regime, the Seebeck coefficient associated to the spin-$\sigma$ channel, $S_{\sigma}$, that appears in Eqs.~\eqref{Sc},~\eqref{Ss},~\eqref{S+} and~\eqref{S-} is:
\begin{equation} \label{S-sigma-int}
  S_{\sigma} = -\frac{1}{G_{\sigma}} \int g_{\sigma}(E) \left( -\frac{\partial f_0}{\partial E} \right) \left( \frac{E - \mu_0}{eT_0} \right) dE,
\end{equation}
where
\begin{equation} \label{G-sigma-int}
  G_{\sigma} = \int g_{\sigma}(E) \left( -\frac{\partial f_0}{\partial E} \right) dE,
\end{equation}
is the conductance of the spin-$\sigma$ channel, and $\mu_0$ and $T_0$ are the chemical potential and the temperature of the electrodes in equilibrium, respectively (see Appendix~\ref{linear_response}). Notice that all the microscopic information regarding the MTJ is now encoded in the transmission probability $T_{\sigma}$ and the Fermi-Dirac distribution.



\subsection{Model} \label{model-sec}

\subsubsection{Hamiltonian} \label{H-sec}

\begin{table*}
  \begin{tabular}{lcc}
    \hline
    \hline
    & Rectangular Potential Barrier & Dirac-delta Function Potential Barrier \\
    \hline
    $m(z)$ & $m_1 \theta(-z + z_l) + m_2 \theta(z - z_l) + (m_3 - m_2) \theta(z - z_r)$ & $m_1 \theta(-z + z_2) + m_3 \theta(z - z_2)$ \\
    $\mathcal{V}(z)$ & $V_0 {\cal I} [\theta(z - z_l) - \theta(z - z_r)]$ & $(V_0 d) {\cal I} \delta(z - z_2)$ \\
    ${\bf H}(z)$ & $-\Delta \mathbf{n} \theta (-z + z_l)$ & $-\Delta \mathbf{n} \theta (-z + z_2)$ \\
    ${\cal V_{BR}}$ & $\frac{1}{\hbar} \alpha (p_x \sigma_y - p_y \sigma_x) \delta (z - z_l)$ & $\frac{\alpha}{\hbar}(p_x \sigma_y - p_y \sigma_x) \delta(z - z_2)$ \\
    ${\cal V_D}$ & $-\frac{1}{\hbar^3}\gamma(p_x \sigma_x - p_y \sigma_y) p_z [\theta(z - z_l) - \theta(z - z_r)] p_z$ & $\frac{\bar{\gamma}}{\hbar}(p_x \sigma_x - p_y \sigma_y) \delta(z - z_2)$ \\
    \hline
    \hline
  \end{tabular}
  \caption{The rectangular barrier and delta function models. The subscripts 1, 2 and 3 correspond to the Fe, GaAs and Au layers, respectively, and in the rectangular barrier model $l$ and $r$ refer to the left and right interfaces (Fig.~\ref{fig-models}). The magnetization direction is given by the unit vector $\mathbf{n}$, the exchange energy by the parameter $\Delta$ and $\theta(z)$ is the Heaviside step function. The Bychkov-Rashba coupling strength at the Fe/GaAs interface is given by the parameter $\alpha$, and $\delta(z)$ is the Dirac-delta function. The bulk Dresselhaus coupling strength is $\gamma$, and $\bar{\gamma}$ is the linearized Dresselhaus parameter.~\cite{Matos-Abiague_PRB_2009}}
  \label{models}
\end{table*}

To study essential effects of anisotropic thermopower and Seebeck effects, we use the model system introduced earlier in Refs.~[\onlinecite{Moser_PRL_2007,Matos-Abiague_PRB_2009}] to explain TAMR experiments in Fe/GaAs/Au tunnel junctions. This model allows an analytical calculation of the spin-dependent tunneling transmission probability $T_{\sigma}$ in the presence of SOI and, in our view, can serve as a benchmark for analyzing TAMT and TASS effects.

The structure we model is shown Fig.~\ref{fig-MTJ}. The metallic layers are described as free and independent electron gases~\cite{Ashcroft} in a semi-infinite space (meaning that the only boundaries are the interfaces between the electrodes and the semiconductor layer). A more sophisticated description is given for the GaAs layer, where the extended Kane model 
is used.~\cite{ACTA} This model accounts for the bulk inversion asymmetry of the GaAs semiconductor (due to its zinc-blende crystal structure) and the structure inversion asymmetry of the MTJ. The former asymmetry causes a D-SOI field, while the later one causes a BR-SOI field. The combination of both spin-orbit fields leads to an overall anisotropic SOI with the required $C_{2v}$ symmetry observed in TAMR experiments using a Fe/GaAs/Au MTJ.~\cite{Moser_PRL_2007} {\it Ab initio} calculations have confirmed that the origin of this symmetry is the atomic structure at both interfaces of the MTJ.~\cite{Gmitra_PRL_2013}

An external magnetic field, ${\bf B_{\rm ext}}$, is used to control the magnetization orientation of the ferromagnetic electrode. This occurs in the saturation limit, where the strength of the magnetic field is such that the magnetization is forced to remain parallel to ${\bf B_{\rm ext}}$. However, the Zeeman splittings due to $\mathbf{B}_{\rm ext}$ are negligible compared to the exchange energy in the ferromagnet. Furthermore, the orbital effects due to ${\bf B_{\rm ext}}$ can be safely neglected as long as the magnetization remains in-plane.~\cite{Wimmer_PRB_2009}

\begin{figure}
  \includegraphics[scale=0.35]{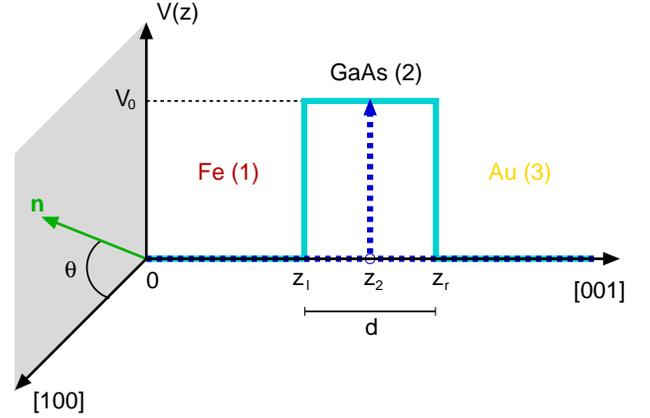}
  \caption{(Color online.) Schematics of the two models used for describing the tunneling barrier (Table~\ref{models}). The solid (dashed) light (dark) blue line represents the rectangular barrier (delta function) model, where $d$ and $V_0$ correspond to the thickness and the height of the GaAs layer, respectively. Additionally, the figure shows the relative orientation $\theta$ of the in-plane magnetization ${\bf n}$ (green arrow) with respect to the $[100]$ crystallographic axis of the GaAs layer.}
  \label{fig-models}
\end{figure}

We use two different models to describe the tunneling barrier, namely, a rectangular potential barrier and a Dirac-delta function potential barrier (Fig.~\ref{fig-models}). Both models have been successful in describing TAMR experiments,~\cite{Moser_PRL_2007} while the later one has been used for describing TAMT experiments as well.~\cite{Naydenova_PRL_2011} We shall next discuss the Hamiltonian for the system and, in the following, the differences between both models will be indicated in Table~\ref{models}.

With all this in mind, the Hamiltonian we use for describing the MTJ is
\begin{equation} \label{Unpert_ham}
\mathcal{H} = \mathcal{T + V + V_Z} + \mathcal{V_{BR} + V_D}.
\end{equation}
Since the MTJ is a heterostructure, the effective mass of the electrons is, in general, different in each layer (although we take it to be constant within each one), meaning the mass becomes position dependent, and so does the kinetic energy operator ${\cal T}$,~\cite{Davies} 
\begin{equation}
\mathcal{T}(z) =  {\bf p} \cdot \left[ \frac{1}{2 m(z)} {\bf p} \right] \mathcal{I},
\end{equation} 
where $\mathbf{p} = -i \hbar \nabla$ is the momentum operator, $m(z)$ is the position dependent effective mass (see Table~\ref{models}) and $\mathcal{I}$ is the unit matrix in spinor space. 
The second term in Eq.~\eqref{Unpert_ham}, $\mathcal{V}$, describes the semiconductor tunneling barrier (see Table~\ref{models}). 
The third term 
, $\mathcal{V_Z}$, accounts for the exchange energy due to the magnetization in the ferromagnetic lead (Stoner model),~\cite{Slonczewski_PRB_1989}
\begin{equation}
\mathcal{V_Z}(z) = \frac{1}{\hbar} \mathbf{H}(z) \cdot {\bf S}.
\end{equation}
$\mathbf{H}(z)$ is the effective exchange field (see Table~\ref{models}), ${\bf S} = (\hbar/2) \boldsymbol{\sigma}$ is the spin operator and $\boldsymbol{\sigma} = (\sigma_x,\sigma_y,\sigma_z)$ are the Pauli matrices. $\mathbf{H}(z)$ is taken to be in-plane, meaning $\mathbf{n} = (\cos \theta, \sin \theta, 0)$, where $\theta$ defines the angle between the magnetization and the GaAs $[100]$ crystallographic axis. In previous experiments with Fe/GaAs/Au MTJs,~\cite{Moser_PRL_2007} the reference axis was taken as the GaAs $[110]$ direction. Therefore, we prefer to express the magnetization direction relative to the [110] axis by introducing the angle shifting $\phi = \theta -\pi/4$.
Finally, the terms ${\cal V_{BR}}$ and ${\cal V_D}$ in Eq.~\eqref{Unpert_ham} account for the BR-SOI and the D-SOI, respectively (see Table~\ref{models}). Previous calculations have shown that the BR-SOI is relevant mainly at the Fe/GaAs interface.~\cite{Matos-Abiague_PRB_2009} Thus, in our calculations the BR-SOI corresponding to the GaAs/Au interface is not considered.

\subsubsection{Tunneling states} \label{PS-sec}

The wave functions describing the conduction electrons are obtained by solving the stationary Pauli-Schr\"odinger equation
\begin{equation} \label{PS_eqn}
 {\cal H} \boldsymbol{\psi}(\mathbf{R}) = E \boldsymbol{\psi}(\mathbf{R}),
\end{equation}
for the Hamiltonian given in Eq.~\eqref{Unpert_ham}, where
\begin{equation}
\boldsymbol{\psi}(\mathbf{R}) = \left( \begin{array}{c} \psi_{\uparrow}(\mathbf{R}) \\ \psi_{\downarrow}(\mathbf{R}) \end{array} \right),
\end{equation}
is a spinor which components correspond to the wave functions for spin-up and spin-down electrons, respectively, and $\mathbf{R} = (x, y, z)$. The solutions of Eq.~\eqref{PS_eqn} are the eigenenergies, $E$, and eigenstates of the system. Since the transversal modes, $\mathbf{p}_{\parallel}$, are conserved,
\begin{equation} \label{ansatz}
  \psi_{\sigma} (\mathbf{R}) = e^{i \mathbf{k_{\parallel} \cdot r}} \varphi_{\sigma}(z),
\end{equation}
where $\varphi_{\sigma}(z)$ is a solution of the Pauli-Schr\"odinger equation for the longitudinal modes, $\mathbf{r} = (x,y)$, $\mathbf{k}_{\parallel} = (k_{x}, k_{y})$ and ${\bf p}_{\|} = \hbar {\bf k}_{\|}$. Notice that the longitudinal (out-of-plane) modes have decoupled from the transverse (in-plane) modes. Replacing Eq.~\eqref{ansatz} in Eq.~\eqref{PS_eqn} allows to solve analytically the resulting one-dimensional stationary Pauli-Schr\"odinger equation (see appendices~\ref{RPB-model} and~\ref{DPB-model}). We find that
\begin{subequations}
\begin{eqnarray}
\boldsymbol{\varphi}_{i}(z) & = & \frac{1}{\sqrt{k_{1\sigma}}} \,  e^{i k_{1\sigma}z} \boldsymbol{\nu}_{\sigma}, \label{phi_i} \\
\boldsymbol{\varphi}_{t}(z) & = & \left( t_{\sigma, \sigma} \boldsymbol{\nu}_{\sigma} + t_{\bar{\sigma}, \sigma} \boldsymbol{\nu}_{\bar{\sigma}} \right) e^{i k_3z}, \label{phi_t}
\end{eqnarray}
\end{subequations}
where
\begin{subequations}
\begin{eqnarray} \label{wave-vectors-electrodes}
  k_{1\sigma} & = & \sqrt{\frac{2m_1}{\hbar^2} \left( E + \sigma \frac{\Delta}{2} \right) - k_{\|}^2}, \\
  k_3 & = & \sqrt{\frac{2m_3E}{\hbar^2} - k_{\|}^2},
\end{eqnarray}
\end{subequations}
and
%
\begin{equation} \label{eigenvectors-electrodes}
  \boldsymbol{\nu}_{\sigma} = \frac{1}{\sqrt{2}} \left( \begin{array}{c} 1 \\ \sigma e^{i\theta} \end{array} \right), \\
\end{equation}
where $\sigma = \uparrow (1), \downarrow (-1)$. The subscripts $i$ and $t$ stand for the incident and the transmitted wave functions, respectively. The coefficient $t_{\sigma, \sigma}$/$t_{\bar{\sigma}, \sigma}$ represents the transmission probability amplitude for a tunneling process in which the electron spin is preserved/flipped. These amplitudes are computed analytically by solving the set of linear equations obtained when imposing the boundary conditions to the wave functions~\eqref{phi_i} and~\eqref{phi_t}. The expressions obtained for $t_{\sigma, \sigma}$/$t_{\bar{\sigma}, \sigma}$ within the rectangular barrier and the delta function models are given in Eq.~\eqref{trans_amp_RPB}/\eqref{trans_amp_RPB-sf} (Appendix~\ref{RPB-model}) and Eq.~\eqref{trans_amp_DPB}/\eqref{trans_amp_DPB-sf} (Appendix~\ref{DPB-model}), respectively.


\subsubsection{Transmission probability} \label{T-sec}

In general,  the transmission probability is defined as
\begin{equation} \label{Tsigma_def}
  T = \frac{J_{z}^{(t)}}{J_{z}^{(i)}},
\end{equation}
where $J_{z}^{(i)}$ and $J_{z}^{(t)}$ are the incident and transmitted probability current densities across the MTJ, respectively. The probability current density is given by the expression
\begin{equation} \label{J}
J_z(z) = \frac{\hbar}{2im(z)} \left[ \boldsymbol{\psi}^{\dagger} \frac{\partial}{\partial z} \boldsymbol{\psi} - \left( \boldsymbol{\psi}^{\dagger} \frac{\partial}{\partial z} \boldsymbol{\psi} \right)^* \right].
\end{equation}
Therefore, using the wave functions for the incident and transmitted electrons computed in the previous section [Eqs.~\eqref{phi_i} and~\eqref{phi_t}], we find that the corresponding current probability densities are
\begin{subequations}
  \begin{eqnarray} \label{crnt_irt}
    J_{z}^{(i)} & = & -\frac{\hbar e}{m_1}, \label{Jzi} \\
%
%
    J_{z}^{(t)} & = & -\frac{\hbar e}{m_3} \left( k_{3\sigma} |t_{\sigma,\sigma}|^2 + k_{3\bar{\sigma}} |t_{\bar{\sigma},\sigma}|^2 \right). \label{Jzt}
  \end{eqnarray}
\end{subequations}
Finally, the transmission probability for a spin-$\sigma$ incoming electron is found by replacing Eqs.~\eqref{Jzi} and~\eqref{Jzt} in Eq.~\eqref{Tsigma_def},
\begin{equation}
T_{\sigma}(E,{\bf k_{\|}}) = \frac{m_1}{m_3} \left( k_{3\sigma} |t_{\sigma, \sigma}|^2 + k_{3\bar{\sigma}} |t_{\bar{\sigma},\sigma}|^2 \right).
\end{equation}


\section{Results} \label{results-sec}

\begin{figure}
  \includegraphics[width=\columnwidth]{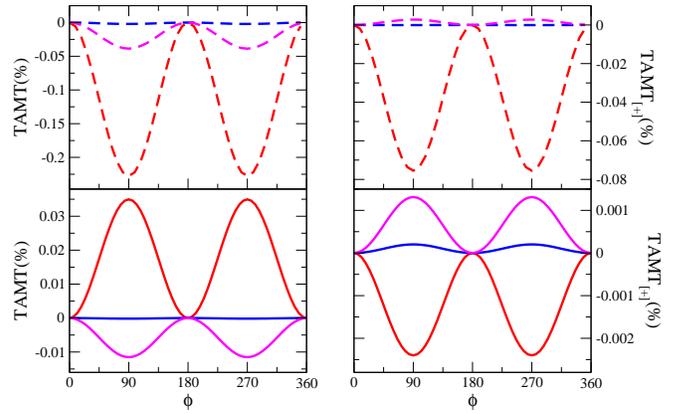}
  \vspace{0.01cm}
  \caption{(Color online.) Tunneling anisotropic magneto-thermopower [Eqs.~\eqref{TAMS} and~\eqref{TAMS+}] dependence on the the angle $\phi$ spanned between the magnetization and the GaAs $[110]$ direction. Solid and dashed lines correspond to the results obtained with the rectangular barrier and the delta function models, respectively (see Table~\ref{models}). The red, blue and magenta lines correspond to the values $\alpha = 42.3\,$eV \AA$^2$ and $\bar{\gamma} = -1.199\,$eV \AA$^2$, $\alpha = -0.6\,$eV \AA$^2$ and $\bar{\gamma} = -3.979\,$eV \AA$^2$, and $\alpha = -17.4\,$eV \AA$^2$ and $\bar{\gamma} = -3.418\,$eV \AA$^2$, respectively, of the Bychkov-Rashba, $\alpha$, and the linearized Dresselhaus, $\bar{\gamma}$, couplings. The values used for the remaining model parameters are: $m_1 = m_3 = m_0$ and $m_2 = 0.067m_0$, where $m_0$ is the bare electron mass, $V_0 = 0.75\,$eV, $d = 80\,$\AA, $\Delta = -3.46\,$eV, and $\gamma = 24\,$eV\AA$^3$.}
  \label{TAMT-fig}
\end{figure}

\begin{figure}
  \includegraphics[width=\columnwidth]{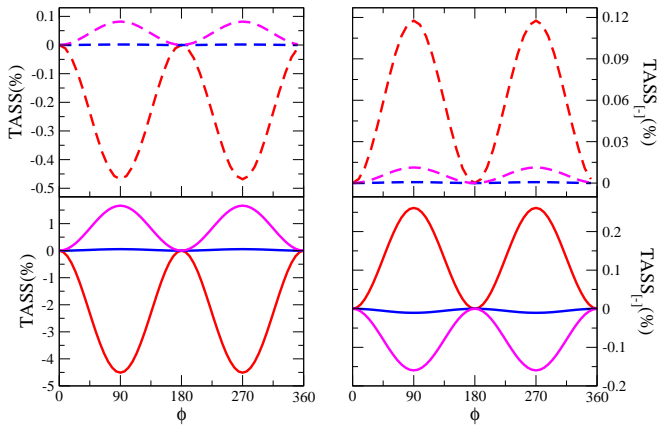}
  \caption{(Color online.) Tunneling anisotropic spin-Seebeck [Eqs.~\eqref{TASS-} and~\eqref{TASSs}] dependence on the the angle $\phi$ spanned between the magnetization and the GaAs [110] crystallographic axis. Idem as in Fig.~\ref{TAMT-fig}.}
  \label{TASS-fig}
\end{figure}

Fig.~\ref{TAMT-fig} shows the TAMT [Eq.~\eqref{TAMS}] and the TAMT$_{[+]}$ [Eq.~\eqref{TAMS+}] calculated with the rectangular barrier model (solid lines) and the delta function model (dashed lines), for different values of the Bychkov-Rashba and the linearized Dresselhaus couplings. Similarly, Fig.~\ref{TASS-fig} shows the TASS$_{[-]}$ [Eq.~\eqref{TASS-}] and the TASS [Eq.~\eqref{TASSs}]. All curves exhibit the expected $C_{2v}$ symmetry, and both models show the same qualitative behavior. However, the quantitative agreement is not so good, namely, they show that the absolute value and the sign of the amplitudes are model dependent.

In the rectangular barrier model, the Bychkov-Rashba coupling $\alpha$ is a phenomenological parameter that is extracted from the experimental data measured for the TAMR in an identical system.~\cite{ACTA,Moser_PRL_2007,Matos-Abiague_PRB_2009} The Dresselhaus spin-orbit coupling in GaAs barrier is $\gamma \approx 24\,$eV\AA$^3$.~\cite{ACTA,Winkler,Teresa_PRL_1999} However, in the delta function model, the linearized Dresselhaus coupling $\bar{\gamma}$ is an additional phenomenological parameter. Thus, in this case $\alpha$ is chosen to be the same as in the rectangular barrier model, while $\bar{\gamma}$ is extracted from the TAMR experimental data. We believe that the main reason for the discrepancy between the rectangular barrier and the delta function models is the fact that the energy dependence of the transmission probability corresponding to the two models is intrinsically different. The transmission probability obtained within the delta model oversimplifies the details and complexity captured by the more accurate rectangular barrier model. Such a discrepancy can be partially solved by properly adjusting the value of the phenomenological $\bar{\gamma}$ parameter in a way that the two models match the experimentally determined TAMR. 
 Indeed, in the linear response regime the conductance is determined by the area comprised by the function $g_{\sigma}(E) (-\partial_E f_0(E))$ [see Eqs.~\eqref{GE} and~\eqref{G-sigma-int}]. 
Thus, by properly choosing the value of the $\bar{\gamma}$ parameter the area resulting from the delta model can be adjusted to that obtained within the rectangular barrier model. This leads to a good agreement between the two models for the TAMR.
 For the calculation of the Seebeck coefficients, however, the situation is different since they are determined by the area of $g_{\sigma}(E) (-\partial f_0(E)) (E - \mu_0)$ [see Eqs.~\eqref{GE} and~\eqref{S-sigma-int}]. Therefore, since the transmission probability obtained within the rectangular and delta models are quite different, it is not possible to simultaneously adjust the values of $G_{\sigma}$ and $S_{\sigma}$ with a single parameter.
 This explains the discrepancies between the two models discussed here when describing the TAMT and TASS in spite of their good agreement previously found for the case of the TAMR.~\cite{Matos-Abiague_PRB_2009}



The maximum values found for the TAMT, TAMT$_{[+]}$, TASS$_{[-]}$ and the TASS are given in Table~\ref{Max-table} for the rectangular barrier and the delta function models. Due to the previous discussion, however, we expect the results obtained with the rectangular barrier model to be more accurate than the ones obtained with the delta model. Furthermore, although the values obtained for the TAMT and TAMT$_{[+]}$ are seemingly small, the corresponding thermovoltages that lead to those values are in principle still measurable. Typically, in these systems the thermovoltages measured are in the range of hundreds of $\mu$V,~\cite{Naydenova_PRL_2011} hence, taking this as a reference, an anisotropy of the order of 10$^{-3}\%$ (the smallest in Table~\ref{Max-table}) can be detected by measuring the thermovoltage with a resolution of the order of nV.



\begin{table}
  \begin{tabular}{lcc}
    \hline \hline
    & Rectangular barrier & Delta barrier \\ \hline
    TAMT & $0.035\%$ & $-0.23\%$ \\
    TAMT$_{[+]}$ & $-0.0024\%$ & $-0.075\%$ \\
    TASS$_{[-]}$ & $0.26\%$ & $0.12\%$ \\
    TASS & $-4.51\%$ & $-0.47\%$ \\ \hline \hline
  \end{tabular}
  \caption{Maximum values obtained for the TAMT [Eq.~\eqref{TAMS}], TAMT$_{[+]}$ [Eq.~\eqref{TAMS+}], TASS$_{[-]}$ [Eq.~\eqref{TASS-}] and the TASS [Eq.~\eqref{TASSs}] within the rectangular barrier model and the delta function model. Due to the discrepancy between both models (see main text), we believe that the results corresponding to the rectangular model are the most accurate.}
  \label{Max-table}
\end{table}


\subsection{Phenomenological model} \label{phe-sec}

Inspired by the phenomenological model developed in Ref.~[\onlinecite{Matos-Abiague_PRB_2009}], we have found that the conductance for the spin-$\sigma$ channel is
\begin{eqnarray} \label{Gsigmaphe}
  G_{\sigma}(E,\phi) & = & G_{\sigma}^{\rm iso}(E) + G_{\sigma}^{\rm aniso}(E) \cos(2 \phi),
\end{eqnarray}
where $G_{\sigma}^{\rm iso}$ and $G_{\sigma}^{\rm aniso}$ are phenomenological parameters,\footnote{The difference between Eq.~\eqref{Gsigmaphe} and the conductance found in Ref.~[\onlinecite{Matos-Abiague_PRB_2009}] is that here it is spin and energy dependent. Apart from this, the derivation is the same. $G_{\sigma}^{\rm iso}$ and $G_{\sigma}^{\rm aniso}$ are given by Eqs.~(52) and~(54) of Ref.~[\onlinecite{Matos-Abiague_PRB_2009}], respectively.} that satisfy that $G_{\sigma}^{\rm iso} \gg G_{\sigma}^{\rm aniso}$. For the special case where no SOI fields are present the anisotropy disappears, meaning $G_{\sigma}^{\rm aniso} = 0$.

Thereupon, taking advantage of Mott's relation for the Seebeck coefficient of the spin-$\sigma$ channel [see Appendix~\ref{Mottslaw} Eq.~\eqref{MottSsigma}] in Eqs.~\eqref{TAMS},~\eqref{TAMS+},~\eqref{TASS-} and~\eqref{TASSs}, respectively, and using Eq.~\eqref{Gsigmaphe} for $G$ and $G_s$, we find that the TAMT, TAMT$_{[+]}$, TASS$_{[-]}$ and the TASS are all proportional to $(1 - \cos2\phi)$. For each case, the corresponding amplitude is a specific function of $G_{\sigma}^{\rm iso}$ and $G_{\sigma}^{\rm aniso}$. Within the phenomenological approach these parameters cannot be computed, meaning that no quantitative predictions can be made. Nevertheless, this description is good enough in order to describe qualitatively the angular dependence found Figs.~\ref{TAMT-fig} and~\ref{TASS-fig}. The quantitative results are obtained using the theory described in the previous section.








\section{Summary} \label{summary-sec}

We have studied thermally induced spin-dependent transport across a three layer MTJ with a single ferromagnetic electrode in the presence of interfacial spin-orbit coupling. Prompted by previous works where TAMR was observed for Fe/GaAs/Au MTJs,~\cite{Moser_PRL_2007} we have shown that a similar anisotropy can be found in the (effective) thermopower and the (effective) spin-Seebeck coefficient, when rotating the magnetization in the ferromagnetic lead with respect to a reference crystallographic axis. This anisotropy is due to the combined effect of the Bychkov-Rashba and the Dresselhaus spin-orbit fields, which posses a characteristic $C_{2v}$ symmetry that appears in the TAMT, TAMT$_{[+]}$, TASS$_{[-]}$ and the TASS, as was the case for the TAMR. The maximum values we have found for the anisotropies of the (effective) thermopower and (effective) spin-Seebeck coefficient are shown in Table~\ref{Max-table}. Finally, since the TAMT effect has recently been experimentally observed in an all-semiconductor MTJ,~\cite{Naydenova_PRL_2011} we believe that the TAMT, TAMT$_{[+]}$, TASS$_{[-]}$ and the TASS effects predicted in this work might also be measurable in similar experiments for Fe/GaAs/Au MTJs. Furthermore, inspired by recent experiments where anisotropic electric spin-injection for ferromagnetic-Silicon interfaces has been observed,~\cite{Sharma_PRB_2012} we hope this work may encourage the observation of anisotropic {\it thermal} spin-injection.


\begin{acknowledgments}
The authors are grateful for the financial support offered by the Deutsche Forshungsgemeinschaft (DFG) via the Priority Program ``Spin Caloric Transport'' (SPP 1538).
\end{acknowledgments}


\appendix

\section{Linear Response} \label{linear_response}

The Fermi-Dirac distribution is
\begin{equation}
  f_{L(R)}(E) = \frac{1}{1 + e^{(E - \mu_{L(R)})/k_BT_{L(R)}}}
\end{equation}
where $\mu_L = \mu_R + eV_{\rm bias}$, $T_L = T_R + \Delta T$, $\mu_L/\mu_R$ and $T_L/T_R$ are the chemical potential and the temperature of the left/right electrode, respectively, and $V_{\rm bias}$ and $\Delta T$ correspond to a bias voltage and a temperature gradient applied to the system, respectively. In the linear response regime,
\begin{equation} \label{flmfr}
  f_L - f_R \simeq \left( -\frac{\partial f_0}{\partial E} \right) \left( eV_{\rm bias} + \frac{\Delta T}{T_0} (E - \mu_0) \right),
\end{equation}
where $\mu_0$ and $T_0$ are the equilibrium values -i.e., when $V_{\rm bias} = 0$  and $\Delta T = 0$. Since
\begin{equation} \label{Seebeck}
  S_{\sigma} = -\frac{1}{G_{\sigma}} \frac{I_{\sigma}}{\Delta T}.
\end{equation}
[see Eq.~\eqref{Isigma}], taking $V_{\rm bias} = 0$ and replacing Eqs.~\eqref{Isint} and~\eqref{flmfr} in Eq.~\eqref{Seebeck} leads to Eq.~\eqref{S-sigma-int} for the Seebeck coefficient for the spin-$\sigma$ channel.


\section{Mott's Law} \label{Mottslaw}

Replacing Eq.~\eqref{flmfr} in Eq.~\eqref{Isint} yields:
\begin{equation}
  I_{\sigma} = \frac{1}{e} \int K_{\sigma}(E) \left( -\frac{\partial f_0}{\partial E} \right) dE,
\end{equation}
where
\begin{equation} \label{KE}
K_{\sigma} (E) := \left( V_{\rm bias} + \frac{E - \mu_0}{eT_0} \Delta T \right) g_{\sigma}(E).
\end{equation}
Using now the Sommerfeld expansion,~\cite{Ashcroft} we find that
\begin{equation} \label{ISommer}
  I_{\sigma} = K_{\sigma} (\mu_0) + \sum_{n = 1}^{\infty} a_n (k_BT_0)^{2n} \left. \frac{d^{2n}}{dE^{2n}} K_{\sigma} \right|_{E = \mu_0},
\end{equation}
where
\begin{equation}
  a_n = 2(2^{2n-1} - 1) \frac{\pi^{2n}}{(2n)!} B_n,
\end{equation}
and $B_n$ are the Bernoulli numbers. Taking $V_{\rm bias} = 0$ and $n = 1$ (which is a good approximation for low temperatures) in Eq.~\eqref{ISommer},  and replacing it together with Eq.~\eqref{KE} in Eq.~\eqref{Seebeck} we obtain
\begin{equation} \label{MottSsigma}
  S_{\sigma} = - \frac{k_B^2}{e} \frac{\pi^2}{3} \left( \left. \frac{d}{dE} \log g_{\sigma}(E) \right|_{E = \mu_0} \right) T_0.
\end{equation}
This expression is the analogous to Mott's Law,~\cite{Mott} but for the spin-$\sigma$ channel. Finally, replacing Eq.~\eqref{MottSsigma} in Eq.~\eqref{Sc} 
we arrive to
\begin{eqnarray}
  S & = & -\frac{k_B^2}{e} \frac{\pi^2}{3} \left( \left. \frac{d}{dE} \log g(E) \right|_{E = \mu_0} \right) T_0, \label{MottSc}
%
\end{eqnarray}
which is nothing else than the well known Mott relation for the thermopower,~\cite{Mott} where $g = g_{\uparrow} + g_{\downarrow}$.


\vspace{0.7cm}

\section{Rectangular potential barrier model} \label{RPB-model}

The one-dimensional stationary Pauli-Schr\"odinger equation for the longitudinal modes is
\begin{equation} \label{1DPS_eqn}
\left( \begin{array}{cc} h(z) &  s(z) \\ s^*(z) & h(z) \end{array} \right) \left( \begin{array}{c} \varphi_{\uparrow}(z) \\ \varphi_{\downarrow}(z) \end{array} \right) = E \left( \begin{array}{c} \varphi_{\uparrow}(z) \\ \varphi_{\downarrow}(z) \end{array} \right),
\end{equation}
where
\begin{widetext}
  \begin{subequations}
    \begin{eqnarray}
      h(z) & = & \frac{\hbar^2 k_{\|}^2}{2m(z)} + p_z \left( \frac{1}{2m(z)} p_z \right) + V_0 [\theta(z - z_l) - \theta(z - z_r)] \label{hRPB} \\
      s(z) & = & -\frac{\Delta}{2} e^{-i\theta} \theta(-z + z_l) - ik_{\|} \alpha e^{-i\theta_{\|}} \delta (z - z_l) - \frac{\gamma(z)}{\hbar^2} k_{\parallel} e^{i\theta_{\|}} p_z^2 \nonumber \\ & + & \sum_{j = l,r} \delta (z - z_j) (\delta_{jl} - \delta_{jr}) \left( \frac{\gamma}{\hbar} \right) ik_{\|} e^{i\theta_{\|}} p_z. \label{sRPB}
    \end{eqnarray}
  \end{subequations}
\end{widetext}
where $\theta_{\|} = \arg (k_y/k_x)$. The stationary solutions of Eq.~\eqref{1DPS_eqn} in the electrodes are the incident and transmitted wave functions given in Eqs.~\eqref{phi_i} and~\eqref{phi_t}, respectively, and the reflected wave function:
\begin{eqnarray} \label{wave_sol_1_3}
\boldsymbol{\varphi}_r(z) & = & r_{\sigma,\sigma} e^{-ik_{1\sigma}z} \boldsymbol{\nu}_{\sigma} + r_{\bar{\sigma},\sigma} e^{-ik_{1\bar{\sigma}}z} \boldsymbol{\nu}_{\bar{\sigma}}
\end{eqnarray}
where $k_{1\sigma}$ and $\boldsymbol{\nu}_{\sigma}$ are given by Eqs.~\eqref{wave-vectors-electrodes} and~\eqref{eigenvectors-electrodes}, respectively, and $r_{\sigma,\sigma}$ and $r_{\bar{\sigma},\sigma}$ are the reflection probability amplitudes, analogous to the transmission probability amplitudes discussed in the main text.
Whereas the solution of Eq.~\eqref{1DPS_eqn} in the tunnel barrier is
\begin{equation}
\boldsymbol{\varphi}_2(z) = \sum_{i = \pm} \left( C_{\sigma,i} e^{k_{2i} z} + D_{\sigma,i} e^{-k_{2i} z} \right) \boldsymbol{\nu}_{2i},
\end{equation}
where
\begin{equation}
k_{2\pm} = \frac{1}{\sqrt{1 \mp \frac{2m_2\gamma k_{\|}}{\hbar^2}}} \sqrt{\frac{2m_2}{\hbar^2} (V_0 - E) + k_{\parallel}^2},
\end{equation}
and
\begin{equation}
\boldsymbol{\nu}_{2\pm} = \frac{1}{\sqrt{2}} \left( \begin{array}{c} 1 \\ \pm e^{-i\theta_{\|}} \end{array} \right).
\end{equation}

Finally, boundary conditions are imposed to the wave functions and their derivatives for computing the reflection and the transmission probability amplitudes. Firstly, the wave functions must be continuous at the interfaces
\begin{eqnarray}
\boldsymbol{\varphi}_1(z_l) & = & \boldsymbol{\varphi}_2(z_l) \\
\boldsymbol{\varphi}_2(z_r) & = & \boldsymbol{\varphi}_3(z_r).
\end{eqnarray}
And secondly, the derivatives of the wave functions must satisfy the following equations:
\begin{widetext}
\begin{subequations}
  \begin{eqnarray}
    \frac{\hbar^2}{2} \left( \frac{1}{m_1} \left. \frac{\partial \varphi_{1\uparrow}}{\partial z} \right|_{z = z_l} - \frac{1}{m_2} \left. \frac{\partial \varphi_{2\uparrow}}{\partial z} \right|_{z = z_l} \right) - ik_{\|} \alpha e^{-i\theta_{\|}} \varphi_{1\downarrow}(z_l) + k_{\|} \gamma e^{i\theta_{\|}} \left. \frac{\partial \varphi_{2\downarrow}}{\partial z} \right|_{z = z_l} & = & 0, \label{lbup} \\
    \frac{\hbar^2}{2} \left( \frac{1}{m_1} \left. \frac{\partial \varphi_{1\downarrow}}{\partial z} \right|_{z = z_l} - \frac{1}{m_2} \left. \frac{\partial \varphi_{2\downarrow}}{\partial z} \right|_{z = z_l} \right) + ik_{\|} \alpha e^{i\theta_{\|}} \varphi_{1\uparrow}(z_l) + k_{\|} \gamma e^{-i\theta_{\|}} \left. \frac{\partial \varphi_{2\uparrow}}{\partial z} \right|_{z = z_l} & = & 0, \label{lbdn}
  \end{eqnarray}
\end{subequations}
for the left barrier and
\begin{subequations}
  \begin{eqnarray}
    \frac{\hbar^2}{2} \left( \frac{1}{m_2} \left. \frac{\partial \varphi_{2\uparrow}}{\partial z} \right|_{z = z_r} - \frac{1}{m_3} \left. \frac{\partial \varphi_{3\uparrow}}{\partial z} \right|_{z = z_r} \right) - k_{\|} \gamma e^{i\theta_{\|}} \left. \frac{\partial \varphi_{2\downarrow}}{\partial z} \right|_{z = z_r} & = & 0, \label{rbup} \\
    \frac{\hbar^2}{2} \left( \frac{1}{m_2} \left. \frac{\partial \varphi_{2\downarrow}}{\partial z} \right|_{z = z_r} - \frac{1}{m_3} \left. \frac{\partial \varphi_{3\downarrow}}{\partial z} \right|_{z = z_r} \right) - k_{\|} \gamma e^{-i\theta_{\|}} \left. \frac{\partial \varphi_{2\uparrow}}{\partial z} \right|_{z = z_r} & = & 0, \label{rbdn}
  \end{eqnarray}
\end{subequations}
\end{widetext}
for the right barrier. The first and second terms in the left hand side of these equations are the so called BenDaniel-Duke boundary conditions,~\cite{BenDaniel_PR_1966} which correspond to the generalization for heterostructures of the requirement that the derivatives of the wave functions should also be continuous at the interfaces. The third and fourth terms in Eqs.~\eqref{lbup} and~\eqref{lbdn} are due to the BR-SOI and D-SOI at the Fe/GaAs interface, respectively. And the third term in Eqs.~\eqref{rbup} and~\eqref{rbdn} is due to the D-SOI at the GaAs/Au interface.

The exact expressions for the transmission amplitudes $t_{\sigma,\sigma}$ and $t_{\sigma,\bar{\sigma}}$ are quite lengthy. However, a simplified analytical expressions for $t_{\sigma,\sigma}$ and $t_{\sigma,\bar{\sigma}}$ can be obtained in the limit $k_{2\pm}d \gg 1$. In such case, one finds the following approximate relations for the tunneling amplitudes:
\begin{subequations}
  \begin{eqnarray}
    t_{\sigma,\sigma} & = & -\frac{D_{\sigma,\sigma}}{D}, \label{trans_amp_RPB} \\
    t_{\sigma,\bar{\sigma}} & = & -\frac{D_{\sigma,\bar{\sigma}}}{D}, \label{trans_amp_RPB-sf}
  \end{eqnarray}
\end{subequations}
where $D = f_-(-)f_+(-) - f_-(+)f_+(+)$, with
\begin{equation}
  f_{\pm}(\lambda) = \frac{d}{2} \left( \frac{m_0}{m_{\pm \lambda}} k_{2\mp \lambda} - i k_3 \right) \left( 1 - \lambda \sigma e^{i(\theta - \theta_{\|})} \right),
\end{equation}
and
\begin{equation}
  \frac{1}{m_{\pm}} = \frac{1}{m_2} \left( 1 \pm \frac{2 m_2 \gamma k_{\|}}{\hbar^2} \right).
\end{equation}
Furthermore, we have
\begin{subequations}
  \begin{eqnarray}
    D_{\sigma,\sigma} & = & \frac{2m_0d}{m_+} k_{2-} f_-(+) g_- - \frac{2m_0d}{m_-} k_{2+} f_-(-) g_+, \nonumber \\ \\
    D_{\sigma,\bar{\sigma}} & = & \frac{2m_0d}{m_+} k_{2-} f_+(-) g_- - \frac{2m_0d}{m_-} k_{2+} f_+(+) g_+. \nonumber \\
  \end{eqnarray}
\end{subequations}
In these equations we introduced the notation
\begin{widetext}
  \begin{equation}
    g_{\pm} = \frac{id\sqrt{k_{1\sigma}} \left[ \left( f_0 \mp h_1 - \frac{m_0d}{m_{\pm}} k_{2\mp} \right) \left( 1 \pm \sigma e^{i(\theta - \theta_{\|})} \right) \mp h_2 \left( 1 \mp \sigma e^{i(\theta - \theta_{\|})} \right) \right] e^{-k_{2\pm}d}}{h_2^2 + \left( f_0 - h_1 - \frac{m_0d}{m_+} k_{2-} \right)\left( f_0 + h_1 - \frac{m_0d}{m_-} k_{2+} \right)},
  \end{equation}
\end{widetext}
where $f_0 = i(k_{1\sigma} + k_{1\bar{\sigma}})d/2$ and
\begin{subequations}
  \begin{eqnarray}
    h_1 & = & \frac{i\sigma d}{2} (k_{1\sigma} - k_{1\bar{\sigma}}) \cos (\theta - \theta_{\|}) - \frac{\alpha k_{\|} Q}{V_0} \sin (2\theta_{\|}), \nonumber \\ \\
    h_2 & = & -\frac{\sigma d}{2} (k_{1\sigma} - k_{1\bar{\sigma}}) \sin (\theta - \theta_{\|}) - i\frac{\alpha k_{\|} Q}{V_0} \cos (2\theta_{\|}), \nonumber \\
  \end{eqnarray}
\end{subequations}
where $Q = 2 m_0 V_0 d/\hbar^2$. These approximate expressions for the tunneling coefficients are valid up to first order in $\exp(-k_{2\pm}d)$. This approximation is appropriate for treating junctions with high and not too thin potential barriers. Taking the Fermi energy as the zero of the energy scale, the height of the barrier is about $V_0 \approx 0.75 \,$eV and $d$ varies from 20 to 200 \AA.

\vspace{0.1cm}

\section{Dirac-delta function barrier model} \label{DPB-model}

The discussion of this model is analogous to the discussion of the rectangular barrier model carried out in the previous section. Therefore, onwards we only highlight the differences between both models. Firstly, Eqs.~\eqref{hRPB} and~\eqref{sRPB} become
\begin{subequations}
  \begin{eqnarray}
    h(z) & = & \frac{\hbar^2 k_{\|}^2}{2m(z)} + p_z \left( \frac{1}{2m(z)} p_z \right) + V_0d \, \delta(z - z_2) \\
    s(z) & = & -\frac{\Delta}{2} e^{-i\theta} \theta(-z + z_l) \nonumber \\ & - & k_{\|} \left( i\alpha e^{-i\theta_{\|}} - \bar{\gamma} e^{i\theta_{\|}} \right) \delta (z - z_2)
  \end{eqnarray}
\end{subequations}
which solutions in the electrodes are the same as in the rectangular barrier model. Secondly, the boundary conditions become:
\begin{equation}
\boldsymbol{\varphi}_1(z_2) = \boldsymbol{\varphi}_3(z_2),
\end{equation}
and
\begin{widetext}
  \begin{eqnarray}
    \frac{\hbar^2}{2} \left( \frac{1}{m_1} \left. \frac{\partial \varphi_{1\uparrow}}{\partial z} \right|_{z = z_2} - \frac{1}{m_3} \left. \frac{\partial \varphi_{3\uparrow}}{\partial z} \right|_{z = z_2} \right) + V_0d \varphi_{1\uparrow}(z_2) - \left[ k_{\|} \left( i\alpha e^{-i\theta_{\|}} - \bar{\gamma} e^{i\theta_{\|}} \right) \right] \varphi_{1\downarrow}(z_2) & = & 0, \\
    \frac{\hbar^2}{2} \left( \frac{1}{m_1} \left. \frac{\partial \varphi_{1\downarrow}}{\partial z} \right|_{z = z_2} - \frac{1}{m_3} \left. \frac{\partial \varphi_{3\downarrow}}{\partial z} \right|_{z = z_2} \right) + V_0d \varphi_{1\downarrow}(z_2) + \left[ k_{\|} \left( i\alpha e^{i\theta_{\|}} + \bar{\gamma} e^{-i\theta_{\|}} \right) \right] \varphi_{1\uparrow}(z_2) & = & 0.
  \end{eqnarray}
\end{widetext}
And finally, the transmission amplitudes are
\begin{subequations}
  \begin{eqnarray}
    t_{\sigma,\sigma} & = & -\frac{8d^2 \sqrt{k_{1\sigma}} (k_{1\bar{\sigma}} + k_3 + iQ)}{\Omega} \nonumber \\ & + & \frac{8id \sqrt{k_{1\sigma}} ({\bf U} \cdot {\bf S}_{\sigma,\sigma})}{\Omega}, \label{trans_amp_DPB} \\
    t_{\sigma,\bar{\sigma}} & = & -\frac{8id \sqrt{k_{1\sigma}} ({\bf U} \cdot {\bf S}_{\sigma,\bar{\sigma}})}{\Omega}, \label{trans_amp_DPB-sf}
  \end{eqnarray}
\end{subequations}
where
\begin{equation}
  \Omega = \Omega_+(-)\Omega_-(+) - \Omega_+(+)\Omega_-(-),
\end{equation}
with
\begin{equation}
  \Omega_{\pm}(\lambda) = d(k_{1,\pm\sigma} + k_3 + iQ) \left( 1 \pm \lambda \right) + 2i({\bf U} \cdot {\bf S}_{\pm \sigma, \lambda \sigma}),
\end{equation}
and $\lambda = \pm 1$. The vectors ${\bf S}_{\sigma,\pm \sigma}$ are given by
\begin{equation}
  {\bf S}_{\sigma,\pm \sigma} = \boldsymbol{\nu}_{\sigma}^{\dagger} \boldsymbol{\sigma} \boldsymbol{\nu}_{\pm\sigma}.
\end{equation}
where $\mathbf{U} = (2 m_0 d/\hbar^2){\bf w}$ with
\begin{equation}
  {\bf w} = \left( -\alpha k_y + \bar{\gamma} k_x, \alpha k_x - \bar{\gamma} k_y, 0 \right).
\end{equation}


%

\bibliography{Seebeck}{}

\end{document}